\documentclass[conference]{IEEEtran}

\usepackage[pdftex]{graphicx}
\usepackage[T1]{fontenc}
\usepackage{color}
\usepackage{varioref}
\usepackage{textcomp}
\usepackage{amsthm}
\usepackage{amsmath}
\usepackage{amssymb}
\usepackage{graphicx}
\usepackage{setspace}
\usepackage{esint}
\usepackage{enumerate}
\usepackage{algorithm}
\usepackage{algpseudocode}
\usepackage{pifont}

\usepackage{enumitem}
\usepackage{enumerate}
\usepackage{enumitem}
\makeatletter

\newcommand{\Rmnum}[1]{\expandafter\@slowromancap\romannumeral #1@}
\makeatother
 
\newtheorem{theorem}{Theorem}

\providecommand{\propositionname}{Proposition}

\ifCLASSINFOpdf
\else
\fi

\usepackage[T1]{fontenc}
\usepackage{etoolbox}

\makeatletter
\patchcmd{\maketitle}{\@fnsymbol}{\@alph}{}{}  % Footnote numbers from symbols to small letters
\makeatother

\title{Decentralized Coded Caching with Distinct \\ Cache Capacities}
\author{\IEEEauthorblockN{Mohammad Mohammadi Amiri, Qianqian Yang, Deniz G\"und\"uz}
\IEEEauthorblockA{Electrical and Electronic Engineering Department, Imperial College London\\
Email: \{m.mohammadi-amiri15, q.yang14, d.gunduz\}@imperial.ac.uk}}
\date{}
\begin{document}

\pagestyle{empty}

\maketitle

\thispagestyle{empty}
\begin{abstract}
\textit{Decentralized coded caching} is studied for a content server with $N$ files, each of size $F$ bits, serving $K$ active users, each equipped with a cache of distinct capacity. It is assumed that the users' caches are filled in advance during the off-peak traffic period without the knowledge of the number of active users, their identities, or the particular demands. User demands are revealed during the peak traffic period, and are served simultaneously through an error-free shared link. A new decentralized coded caching scheme is proposed for this scenario, and it is shown to improve upon the state-of-the-art in terms of the required delivery rate over the shared link, when there are more users in the system than the number of files. Numerical results indicate that the improvement becomes more significant as the cache capacities of the users become more skewed.   
\\ 
\end{abstract}


\section{Introduction}\label{Intro}
The ever-increasing mobile data traffic has imposed a great challenge on the current network architectures. The growing demand is typically addressed by increasing the achievable data rates; however, moving content to the network edge has recently emerged as a promising alternative solution as it reduces both the bandwidth requirements and the delay. The use of edge caching is further motivated by the continuous drop in the cost of memory. In this paper, we consider an extreme form of edge caching, in which contents are stored directly at user terminals in a proactive manner. Proactive caching of popular content during off-peak traffic periods also helps flattening the high temporal variability of traffic.  \cite{DowdyCaching,AlmerothCacing}.

Proactive caching is performed in two phases: The \textit{placement phase} takes place during off-peak traffic hours, when the resources are abundant, and the users' caches are filled by the server without knowing the future user demands. When the user demands are revealed, the \textit{delivery phase} is performed, in which a common message is transmitted from the server to all the users over the shared communication channel. Each user decodes its requested file by combining the bits received in the delivery phase with the contents of its local cache. The goal is to minimize the \textit{delivery rate}, which guarantees that all the user demands are satisfied, independent of the demand combination of the users. 

Research on caching over the past decade has mainly focused on the placement phase in order to identify the most popular contents to be cached locally at user terminals \cite{baev2008approximation, BorstCaching}. Recently, \textit{coded caching} scheme was introduced in \cite{MaddahAliCentralized} for proactive caching, and it is shown that by storing and transmitting coded contents, and designing the placement and delivery phases jointly, it is possible to significantly reduce the delivery rate compared to uncoded caching. 

A \textit{centralized} caching scenario is studied in \cite{MaddahAliCentralized}, in which the number and the identities of the users are known in advance by the server. This allows coordination of the cache contents across the users during both the placement and delivery phases; such that, by carefully placing pieces of contents in user caches, a maximum number of multicasting opportunities are created for tranmission during the delivery phase. Several other recent work has considered centralized coded caching, and the required delivery rate has been further reduced \cite{ZhiChenXOR,MohammadDenizTCom,KaiWanUncodedCaching}.

In practice, however, the number or the identity of active users that will participate in the delivery phase might not be known in advance during the placement phase. In such a scenario, called \textit{decentralized coded caching}, coordination across users is not possible during the placement phase. However, Maddah-Ali and Niesen proposed a scheme that randomly caches parts of each content at each user, and can still exploit multicasting opportunities in the delivery phase, albeit limited compared to the centralized setting \cite{MaddahAliDecentralized}. Decentralized coded caching has been studied in various other settings, e.g., files with different popularities \cite{NiesenNonuniform, JiArXivNonuniform}, and distinct lengths \cite{ZhangDistinctFileSizes}, online caching \cite{PedarsaniOnlineCaching}, etc. 

Most of the existing literature on coded caching assume identical cache sizes across users. Recently, in \cite{WangHeterogenous} decentralized caching to users with heterogeneous cache sizes is studied, and by extending the scheme proposed in \cite{MaddahAliDecentralized} to this scenarios, authors have shown that significant gains can still be obtained compared to uncoded caching. In this paper, we propose a novel decentralized caching algorithm for users with distinct cache capacities. We show that the proposed scheme requires a smaller delivery rate than the one achieved in \cite{WangHeterogenous}. The simulation results illustrate that the improvement in the delivery rate is more significant when the distribution of the cache capacities across users is more skewed.       

The rest of this paper is organized as follows. The system model is introduced in Section \ref{SystemModel}. In Section \ref{s:Results}, we introduce the proposed coded caching scheme, analyze its performance in terms of the delivery rate. The performance of the proposed caching scheme is compared with the state-of-the-art result, and some numerical results are presented in Section \ref{s:Comparison}. We  conclude the paper in Section \ref{Conc}. 

\textit{Notations:} The set of integers $\left\{ 1, ..., K \right\}$ is denoted by $\left[ 1:K \right]$. Notation $\oplus$ illustrates the bitwise XOR operation. For two sets $Q$ and $P$, $Q \backslash P$ is a set including the members of $Q$ and excluding the members of $P$. Notation $\left| {.} \right|$ represents cardinality of a set or length of a file. We use the notation $\bar \oplus$ to represent the bitwise XOR operation between binary sequences with different lengths. The arguments of $\bar \oplus$ are first zero-padded to have the same length as the longest argument, and then they are bitwise XOR-ed.  

\section{System Model}\label{SystemModel}
A server with $N$ independent $F$-bit files, $W_1, ..., W_N$, is considered, where each file is assumed to be uniformly distributed over $\left[ 1:2^F \right]$. There are $K$ active users, $U_1, ..., U_K$, where user $U_k$ is equipped with a cache of capacity $M_{k}F$ bits, with $M_k \le N$, $\forall k$. We denote the cache capacities by vector $\mu \triangleq (M_1,\ldots,M_K)$. Let $Z_k$ denote the contents of $U_k$'s cache at the end of the \textit{placement phase}. Unlike in centralized coded caching \cite{MaddahAliCentralized}, cache contents are independent of the number of users, their identities, or the user requests. User requests are revealed after the placement phase, where $d_k \in \left[ 1:N \right]$ denotes the file requested by user $U_k$, $k=1, \ldots, K$. User demands are served simultaneously through an error-free shared link. Let $X$ denote the $RF$-bit message transmitted over the shared link by the server to enable each user $U_k$ to decode its requested file $W_{d_k}$, together with its local cache content. Our goal is to characterize the minimum rate $R(\mu)$; such that, each user can decode its desired file with arbitrarily small probability of error, independent of the particular demand combination.

%It is expected that higher cache capacity of each user leads to a reduction in the delivery rate. Hence, there should be a trade-off between the delivery rate and the set of cache capacities $\mu  = \left\{ {{M_1},...,{M_K}} \right\}$, denoted by $R(\mu)$. 

\section{Decentralized Coded Caching}\label{s:Results}

We first illustrate our decentralized coded caching scheme on the following example.

\theoremstyle{definition}
\newtheorem{exmp}{Example}

\begin{exmp}\label{DecDistCacheSizes}
Consider a caching system with $N = 2$ files $W_1$ and $W_2$, and $K = 4$ users. It is assumed that the cache capacity of user $U_k$ is ${M_k} = {\left( {1/2} \right)^{4 - k}}M$, $\forall k \in \left[ 1:4 \right]$. 

In the placement phase, user $U_k$ caches a random $M_{k}F/2$ bits of each file independently. Since there are $N=2$ files in the database, a total of $M_{k}F$ bits are cached by user $U_k$. 

When $N<K$, it can be shown that the worst-case user demands happens when $N$ users with the smallest cache capacities have different requests. For this particular example, we have $M_1 \le \cdots \le M_4$, and the worst-case happens when users $U_1$ and $U_2$ request distinct files. Hence, we can assume the worst-case demand combination of $d_k = 1$, if $k=1,3$, and $d_k = 2$, otherwise.

The contents served in the delivery phase are divided into three distinct parts, where $X_i$ is delivered in part $i$, for $i=1, 2, 3$. Thus, the common message is $X=\left( X_1, X_2, X_3 \right)$. We further divide the message $X_2$ into three pieces $X_2^1$, $X_2^2$, and $X_2^3$. Below, we explain the purpose of each part in detail. 

\begin{enumerate}[label=\bfseries Part \arabic*:,align=left]
\item In the first part of the delivery phase, the bits of each requested file which have not been cached by any user are directly delivered by the server. The following contents are delivered in this part. $X_1=\left( W_{1,\left\{ \emptyset  \right\}}, W_{2,\left\{ \emptyset  \right\}} \right)$.   
\item The bits of the file requested by a user having been cached by another user are transmitted in the second part of the delivery phase. The server first delivers each user the bits of its requested file which are in the cache of one user with the same request. Then, each user receives the bits of its requested file which are in the cache of a single user with different request. By delivering the following contents, user $U_k$ can obtain the bits of file $W_{d_k}$ having been cached in user $U_l$, for $k,l \in \left[1:4\right]$, such that $l \ne k$. $X_2^1$ $=$ $\left( W_{1,\left\{ 3 \right\}} \right.$ $\bar \oplus$ $W_{1,\left\{ 1 \right\}}$, $W_{2,\left\{ 4 \right\}}$ $\bar \oplus$ $\left. W_{2,\left\{ 2 \right\}} \right)$, $X_2^2$ $=$ $\left( W_{1,\left\{ 4 \right\}} \right.$ $\bar \oplus$ $W_{1,\left\{ 2 \right\}}$, $W_{2,\left\{ 3 \right\}}$ $\bar \oplus$ $\left. W_{2,\left\{ 1 \right\}} \right)$, $X_2^3$ $=$ $\left( W_{1,\left\{ 2 \right\}} \right.$ $\bar \oplus$ $\left. W_{2,\left\{ 1 \right\}} \right)$.      
\item In the last part, the server delivers the users the bits of their requested files which have been cached by more than one another user. Accordingly, each user $U_k$, $\forall k \in \left[ 1:4 \right]$, can obtain all the bits of file $W_{d_k}$ which are in the cache of users in any set $S \subset \left[ {1:4} \right]\backslash \left\{ k \right\}$, where $\left| S \right| \ge 2$, after receiving the following contents. $X_3$ $=$ $\left( W_{1,\left\{ 2,3 \right\}} \right.$ $\bar \oplus$ $W_{2,\left\{ 1,3 \right\}}$ $\bar \oplus$ $W_{1,\left\{ 1,2 \right\}}$, $W_{1,\left\{ 2,4 \right\}}$ $\bar \oplus$ $W_{2,\left\{ 1,4 \right\}}$ $\bar \oplus$ $W_{2,\left\{ 1,2 \right\}}$, $W_{1,\left\{ 3,4 \right\}}$ $\bar \oplus$ $W_{1,\left\{ 1,4 \right\}}$ $\bar \oplus$ $W_{2,\left\{ 1,3 \right\}}$, $W_{2,\left\{ 3,4 \right\}}$ $\bar \oplus$ $W_{1,\left\{ 2,4 \right\}}$ $\bar \oplus$ $W_{2,\left\{ 2,3 \right\}}$, $W_{1,\left\{ 2,3,4 \right\}}$ $\bar \oplus$ $W_{2,\left\{ 1,3,4 \right\}}$ $\bar \oplus$ $W_{1,\left\{ 1,2,4 \right\}}$ $\bar \oplus$ $\left. W_{2,\left\{ 1,2,3 \right\}} \right)$.       
\end{enumerate} 
After these parts, each user can decode all the missing bits of its desired file. To find the delivery rate, we first note that, by the law of large number, the length of the subfile $W_{k,V}$, for any set $V \subset \left[ {1:K} \right]$, is approximately given by
\begin{equation}\label{SizeSubfileExample} \left| {{W_{k,V}}} \right| \approx \prod\limits_{i \in V} {\left( {\frac{{{M_i}}}{2}} \right)} \prod\limits_{j \in \left[ {1:4} \right]\backslash V} {\left( {1 - \frac{{{M_j}}}{2}} \right)} F, \quad \forall k \in \left[ {1:K} \right].
\end{equation} 
For the example under consideration, when $M = 1$, i.e., $\mu  = \left\{ {1/8,1/4,1/2,1} \right\}$, the delivery rate is $1.758$, while the delivery rate of the scheme proposed in \cite{WangHeterogenous} for this setting is $2.681$. Hence, the proposed scheme provides $34.43\%$ reduction in the delivery rate compared to the state-of-the-art result for this example.  
\qed
\end{exmp}

\subsection{Placement Phase}
Since the active users are not known in advance in the decentralized setting, cache contents cannot be coordinated among the users. Similarly to the placement phases of the the decentralized coded schemes in the literature \cite{MaddahAliDecentralized, WangHeterogenous}, user $U_k$ caches a random $M_{k}F/N$ bits of each file independently, for $k=1, ..., K$. Since $N$ files are hosted in the database, a total of $M_{k}F$ bits are cached by each user $U_k$, and hence, the corresponding cache-capacity constraint is satisfied. 

For any set $V \subset \left[ {1:K} \right]$, let $W_{i,V}$ represent the bits of file $W_i$ that have been \textit{exclusively} cached by the users in set $V$ at the end of the placement phase, i.e., $W_{i,V} \subset Z_k$, $\forall k \in V$, and $W_{i,V} \cap  Z_k = \emptyset$, $\forall k \in \left[ 1:K \right] \backslash V$.

\subsection{Delivery Phase}
User demands are revealed at the beginning of the delivery phase. Without loss of generality, we re-label the users such that the first $K_1$ users, referred to as group $G_1$, have the same request $W_1$, the next $K_2$ users, group $G_2$, request file $W_2$, and so on so forth. For notational convenience, we define ${S_i} \buildrel \Delta \over = \sum\limits_{l = 1}^i {{K_l}}$. Therefore, the user demands are as follows:
\begin{equation}\label{DemandsGeneralCase} d_k = i,\quad \mbox{for $i=1, ..., N,$ and $k=S_{i-1} + 1, ..., S_i$},
\end{equation} 
where we set $S_0 = 0$. We further order the users within a group according to their cache sizes, and assume, without loss of generality, that ${M_{{S_{i - 1}} + 1}} \le {M_{{S_{i - 1}} + 2}} \le \cdots \le {M_{{S_i}}}$, for $i= 1, \ldots ,N$.

The proposed delivery phase is presented in Algorithm \ref{DeliveryHeterogenous}. For any general demand combination described above, the delivery phase presented in Algorithm \ref{DeliveryHeterogenous} contains two procedures CODED DELIVERY and RANDOM DELIVERY, and in each case the server chooses the one with the smaller delivery rate. Below, we explain these two procedures in detail.  

The CODED DELIVERY procedure includes three distinct parts, where the content delivered in part $i$ is denoted by $X_i$, $i=1,2,3$, and the common message $X=\left( X_1, X_2, X_3 \right)$ is sent to all the users during the delivery phase. The message transmitted in part 2, $X_2$, is further divided into three pieces $X_2^1$, $X_2^2$, and $X_2^3$, i.e., $X_2=\left( X_2^1, X_2^2, X_2^3 \right)$. Based on the aforementioned placement phase, the main motivation of the CODED DELIVERY procedure is to enable each user to recover the missing bits of its requested file which have been cached by $i$ other users, $\forall i \in \left\{ 0, ..., K-1 \right\}$.  

\begin{algorithm}[t]
\caption{Coded Delivery Phase}
\label{DeliveryHeterogenous}
\begin{algorithmic}[1]
\Statex
\Procedure {Coded Delivery}{}
\State{\textbf{Part 1}: Delivering bits that are not in the cache of any user}
\For {$i = 1, 2, \ldots, N$}
\State{$X_1=\left( W_{{d_{S_{i-1} + 1}},\left\{ \emptyset \right\}}  \right)$}
\EndFor
\Statex
\State{\textbf{Part 2}: Delivering bits that are in the cache of only one user}
\State{$X_2^1=\left( {\bigcup\limits_{i = 1}^N {\bigcup\limits_{k = {S_{i - 1}} + 1}^{{S_i} - 1} {\left( {{W_{i,\left\{ k \right\}}} \bar \oplus {W_{i,\left\{ k + 1 \right\}}}} \right)} } } \right)$}
\State{$X_2^2=\bigcup\limits_{i = 1}^{N - 1} \bigcup\limits_{j = i + 1}^N \left( \bigcup\limits_{k = {S_{j - 1}} + 1}^{{S_j} - 1} {\left( {{W_{i,\left\{ k \right\}}} \bar \oplus {W_{i,\left\{ {k + 1} \right\}}}} \right)} ,\qquad \qquad \qquad \qquad \qquad \qquad \qquad \right.$ $ \left. \qquad \qquad \qquad \qquad \qquad \bigcup\limits_{k = {S_{i - 1}} + 1}^{{S_i} - 1} {\left( {{W_{j,\left\{ k \right\}}} \bar \oplus {W_{j,\left\{ {k + 1} \right\}}}} \right)} \right)$} 
\State{$X_2^3=\left( \bigcup\limits_{i = 1}^{N - 1} \bigcup\limits_{j = i + 1}^N W_{i,\left\{S_{j-1}+1\right\}} \bar \oplus W_{j,\left\{S_{i-1}+1\right\}}   \right)$}
\Statex
\State{\textbf{Part 3}: Delivering bits that are in the cache of more than one user}
\For{$i = 1, 2,  \ldots, K - 2$ }
\For{$j = 2, 3, \ldots, K - i$ }
\For{$V \subset \left[ {i + 1:K} \right]: \left| V \right| = j$ }
\State{${X_3} = \left( {\left( {\mathop {\bar  \oplus }\limits_{v \in V} {W_{{d_v},\left\{ {V,i} \right\}\backslash \left\{ v \right\}}}} \right)\bar  \oplus {W_{{d_i},V}}} \right)$}
\EndFor
\EndFor
\EndFor
\EndProcedure
\Statex
\Procedure {Random Delivery}{}
\For{$i = 1, 2, \ldots, N$}
\State {server sends enough random linear combinations of the bits of file $W_{i}$ to enable the users demanding it to dedcode it.}
\EndFor
\EndProcedure
\end{algorithmic}
\end{algorithm}

In Part 1 of the this procedure, each user receives the bits of its requested file which have not been cached by any user. 

The purpose of Part 2 is to enable each user to obtain all the missing bits of its request that have been cached by another single user. First, consider the message $X_2^1$. For $i=1, ..., N$, each user $k \in \left[ S_{i-1}+1:S_{i} \right]$ (i.e., $U_k \in G_i$), has access to bits $W_{i,\{k\}}$ locally in its cache, and with $X_2^1$ it can decode all the pieces $W_{i,\left\{ l \right\}}$, $\forall l \in \left[ S_{i-1}+1:S_{i} \right]$, i.e., the bits of its demand $W_i$, which are in the cache of another user in the same group, and no other user. Delivering the messages $X_2^2$ and $X_2^3$ together helps the users to decode the bits of their requested files having been cached by a single user in other groups; that is, after receiving $\bigcup\limits_{k = {S_{j - 1}} + 1}^{{S_j} - 1} \left( W_{i,\left\{ k \right\}} \bar \oplus W_{i,\left\{ {k + 1} \right\}}\right)$, $\bigcup\limits_{k = {S_{i - 1}} + 1}^{{S_i} - 1} \left( W_{j,\left\{ k \right\}} \bar \oplus W_{j,\left\{ {k + 1} \right\}} \right)$, and $W_{i,\left\{S_{j-1}+1\right\}}$ $\bar \oplus$ $W_{j,\left\{S_{i-1}+1\right\}}$, the users in both groups $G_i$ and $G_j$ can obtain the missing bits of their requested files that have been cached by a user in another group, for $i=1, ..., N-1$ and $j=i+1, ..., N$ (and no other user). Note that, having received $X_2^2$, the third message $W_{i,\left\{S_{j-1}+1\right\}}$ $\bar \oplus$ $W_{j,\left\{S_{i-1}+1\right\}}$ is the smallest number of bits (based on the assumption ${M_{{S_{l - 1}} + 1}} \le {M_{{S_{l - 1}} + 2}} \le \cdots \le {M_{{S_l}}}$, $\forall l \in \left[ 1:N \right]$) that enable all the users in both groups $G_i$ and $G_j$ to obtain the missing bits of their desired files that are in the cache of users in the other group, for $i=1, ..., N-1$ and $j=i+1, ..., N$.   

Part 3 of our algorithm is the same as the delivery phase proposed in \cite[Algorithm 2]{WangHeterogenous}, and it is performed to send the users the missing bits of their requests that have been cached by more than one user. 

Finally, in the RANDOM DELIVERY procedure, as in the DELIVERY procedure of \cite{MaddahAliDecentralized}, the server transmits enough random linear combinations of the bits of file $W_i$ to the users in group $G_i$ to make sure they all can decode the file, for $i=1, \ldots, N$.

\subsection{Delivery Rate Analysis}
In the following, we evaluate the delivery rate of the proposed caching scheme for the worst-case user demands. Consider first the case $N \ge K$. It can be argued in this case that the worst-case user demands happens if each file is requested by at most one user. Hence, by re-ordering the users, for the worst-case user demands, we have $K_i = 1,$ for $1 \le i \le N$, and $K_i = 0,$ otherwise. In this case, it can be shown that the CODED DELIVERY procedure requires a lower delivery rate than the RANDOM DELIVERY procedure; hence, the server uses the former. In this case, it is possible to simplify the CODED DELIVERY procedure such that, only message $X_2^3$ is transmitted in Part 2, when $N \ge K$, i.e., $X_2 = X_3^2$. The corresponding common message, $X=\left( X_1, X_2^3, X_3 \right)$, transmitted over the CODED DELIVERY procedure, reduced to the delivery phase of \cite[Algorithm 2]{WangHeterogenous}. Thus, the proposed scheme achieves the same delivery rate as \cite[Algorithm 2]{WangHeterogenous} when $N \ge K$. 

Next, we consider the case $N<K$. It is possible to show that the worst-case user demands in this case happens when $N$ users with the smallest cache capacities all request different files, i.e., they end up in different groups. The delivery rate of the proposed delivery phase when $N<K$ is presented in the following theorem. The proof of the worst-case demand distribution as well as Theorem \ref{TheDelRateDecDistCacheSizes} are skipped due to space limitations; however, they can be found in the longer version of the paper in \cite{Asilomar16}. 

\begin{theorem}\label{TheDelRateDecDistCacheSizes}
In a decentralized caching system with $N$ files in the database, each of size $F$ bits, and $K$ users with cache capacities $\mu  = \left\{ {{M_1},...,{M_K}} \right\}$, such that $M_1 \le M_2 \le \cdots \le M_K$, the following delivery rate-cache capacity trade-off is achievable when $N<K$:
\begin{align}\label{OurDeliveryRateHeterogenous} R_c\left( \mu  \right) = & \min \left\{ \sum\limits_{i = 1}^K {\left[ {\prod\limits_{j = 1}^i {\left( {1 - \frac{{{M_j}}}{N}} \right)} } \right]} \right.  \nonumber\\
& \quad \;\; \left.- \Delta {R_1}\left( \mu  \right) - \Delta {R_2}\left( \mu  \right),\sum\limits_{i = 1}^N \left( {1 - \frac{{{M_i}}}{N}} \right)  \right\},
\end{align}
where
\begin{subequations}
\label{DeltaR}
\begin{align}\label{DeltaRone}
\Delta {R_1}\left( \mu  \right) =& \left( {K - N} \right)\prod\limits_{l = 1}^K {\left( {1 - \frac{{{M_l}}}{N}} \right)},\\
\Delta {R_2}\left( \mu  \right) =& \left[ {\sum\limits_{k = 1}^{K - N} {\left( {\frac{ (k-1) M_{k + N} }{{N - {M_{k + N}}}}} \right)} } \right]\prod\limits_{l = 1}^K {\left( {1 - \frac{{{M_l}}}{N}} \right)}.
\label{DeltaRtwo}
\end{align}
\end{subequations}
\end{theorem}

\begin{figure}[!t]
\centering
\includegraphics[scale=0.56]{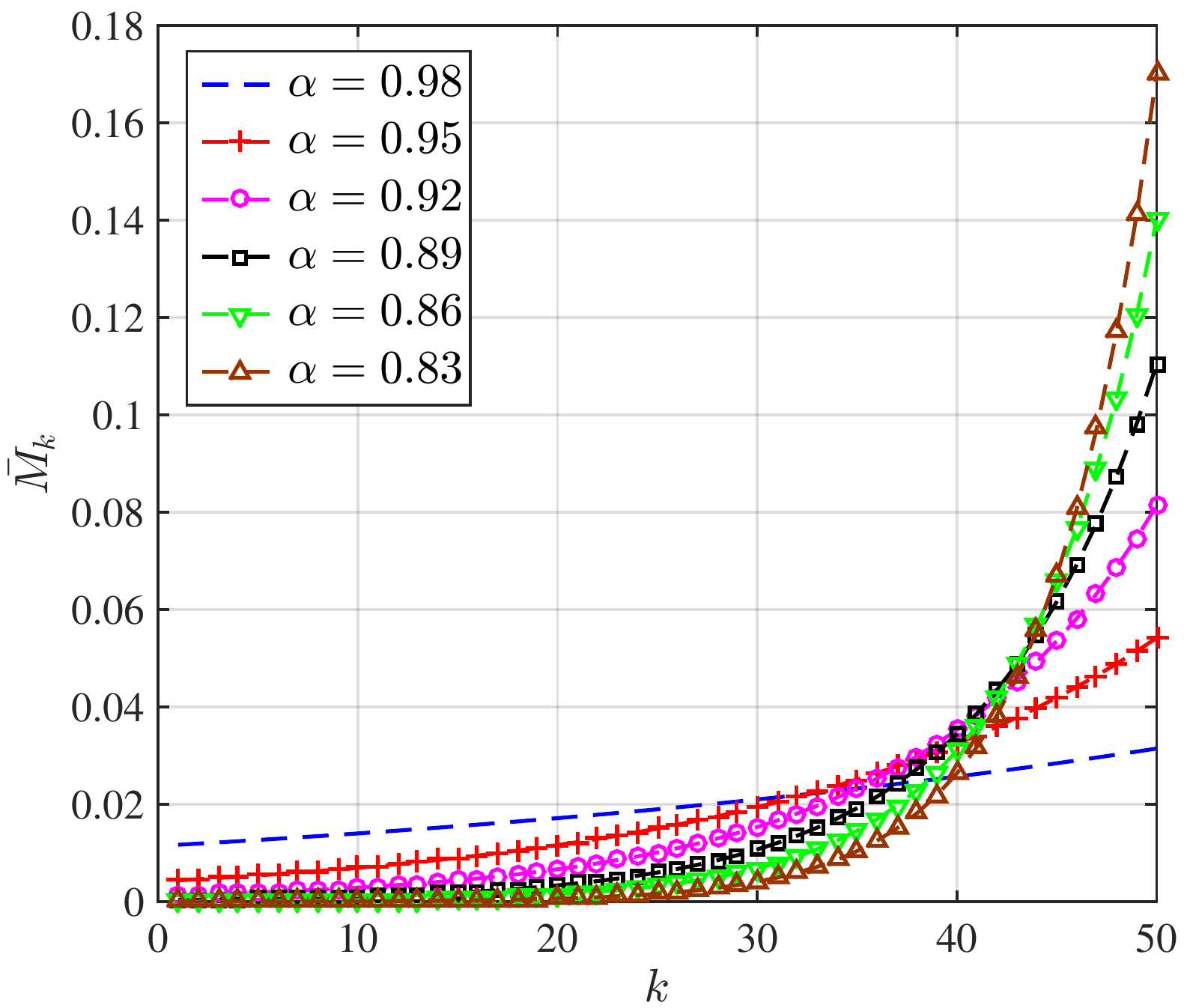}
\caption{Illustration of cache capacity distribution normalized by $\sum\limits_{k = 1}^K {{M_k}}$ for different $\alpha$ values, and $K=50$. The $x$-axis corresponds to the user index $k$.} 
\label{CacheCapacitiesDistribution}
\end{figure}

\section{Comparison with the State-of-the-Art and Numerical Results}\label{s:Comparison}

In this section, the proposed caching scheme is compared with the scheme proposed in \cite{WangHeterogenous} both analytically and numerically. We note that, although the scheme presented in \cite{WangHeterogenous} is for $N \ge K$, it can also be applied to the case $N < K$, and the same delivery rate as \cite[Theorem 2]{WangHeterogenous}, denoted here by $R_b(\mu)$, can be achieved. Hence, in the following, when we refer to the scheme stated in \cite[Algorithm 2]{WangHeterogenous} for $N < K$, we consider its generalization to this scenario. When $N<K$, according to \cite[Theorem 2]{WangHeterogenous} and \eqref{OurDeliveryRateHeterogenous}, we have 
\newcommand\myfirstinequality{\mathrel{\overset{\makebox[0pt]{\mbox{\normalfont\tiny\sffamily (a)}}}{>}}}
\begin{equation}\label{DeliveryRateHeterogenousComparison} {R_b}\left( \mu  \right) - {R_c}\left( \mu  \right) \ge \Delta {R_1}\left( \mu  \right) + \Delta {R_2}\left( \mu  \right) \myfirstinequality 0.
\end{equation}
The inequality (a) holds as long as $N < K$. Therefore, when the number of files in the database is smaller than the number of active users in the delivery phase, the proposed coded caching scheme requires a smaller delivery rate than the one presented in \cite{WangHeterogenous}.

For the numerical results, we consider an exponential cache distribution among users, such that the cache capacity of user $U_k$ is given by 
\begin{align}
M_k = {\alpha ^{K - k}}M,
\end{align}
where $0 \le \alpha \le 1$, for $k=1, \ldots, K$, and $M$ denote the maximum cache capacity in the system. Thus, we have $\mu  = \left\{ {{\alpha ^{K - 1}}M,{\alpha ^{K - 2}}M, \ldots, M} \right\}$, such that $M_1 \le M_2 \le \cdots \le M_K$. The distribution of cache capacities normalized by $\sum\limits_{k = 1}^K {{M_k}}$, i.e., $M_k/\sum\limits_{k = 1}^K {{M_k}}$ denoted by ${\bar M}_K$, $\forall k \in \left[1:K \right]$, is demonstrated in Fig. \ref{CacheCapacitiesDistribution} for different values of $\alpha$, when $K=50$. Observe that, the smaller the value of $\alpha$, the more skewed the cache capacity distribution across users become. In the special case of $\alpha=1$, we obtain the homogeneous cache capacity model studied in \cite{MaddahAliDecentralized}.

\begin{figure}[!t]
\centering
\includegraphics[scale=0.485]{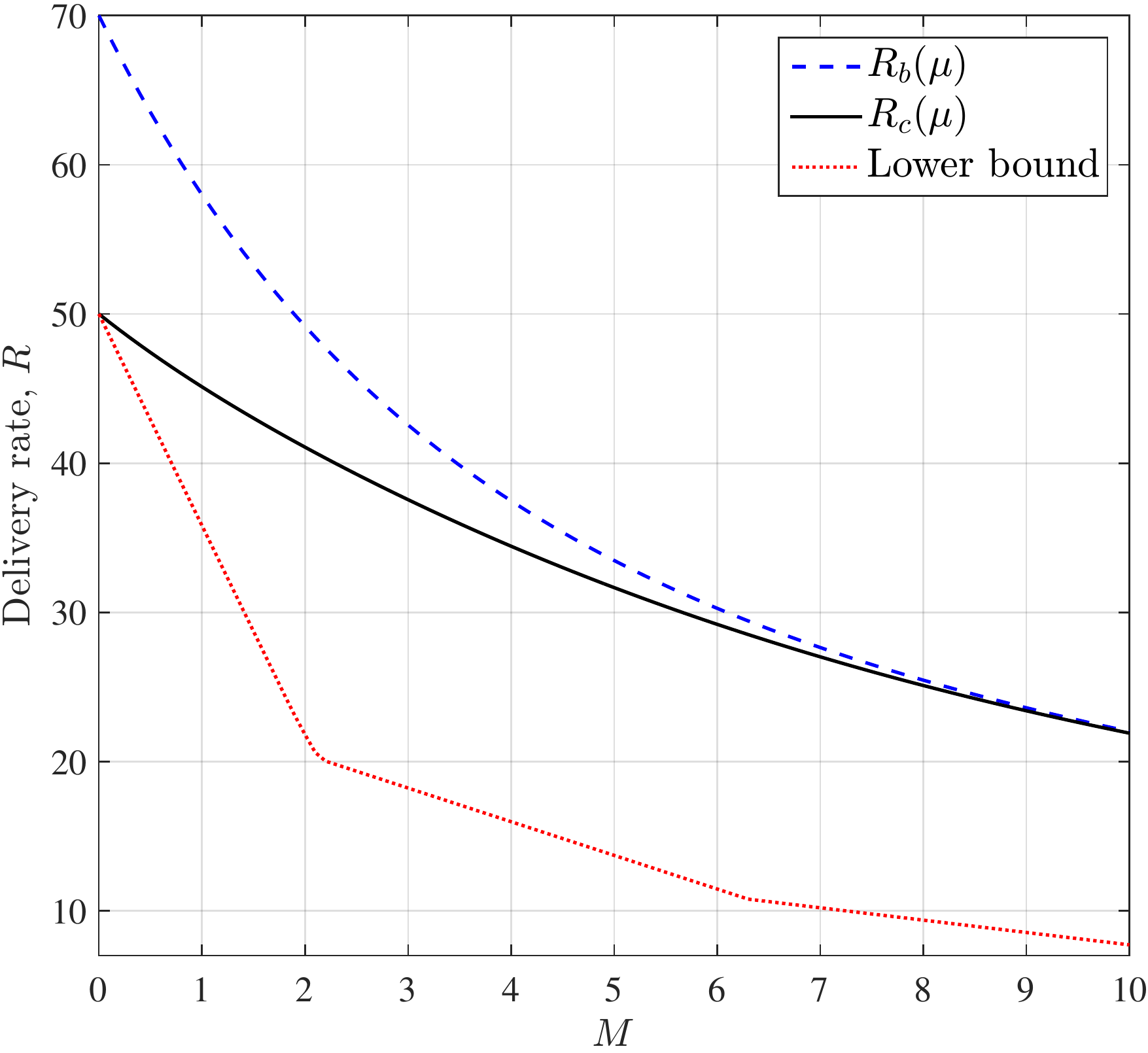}
\caption{Delivery rate versus $M$, where the cache capacity of user $k$ is $M_k = {\alpha ^{K - k}}M$, $k=1, \ldots, K$, when $\alpha=0.97$, $N=50$, and $K=70$.} 
\label{N50K70}
\end{figure}

In Fig. \ref{N50K70}, the delivery rate of the proposed scheme, $R_c(\mu)$, is compared with that of the coded scheme proposed in \cite{WangHeterogenous}, i.e., $R_b(\mu)$, when $N=50$, $K=70$, and $\alpha = 0.97$. The delivery rate is plotted in this figure versus the largest cache capacity in the system, $M$. As expected the performance improves, i.e., the delivery rate reduces as $M$ increases. We also clearly observe that the proposed scheme outperforms the scheme presented in \cite{WangHeterogenous}. The improvement is particularly significant for lower values of $M$. The cut-set lower bound for this setting is also included in the figure. Although the delivery rate of the proposed scheme approaches the lower bound for relatively small values of $M$, there is still a gap for large values of $M$, which may as well be due to the looseness of the lower bound.  

\begin{figure}[!t]
\centering
\includegraphics[scale=0.49]{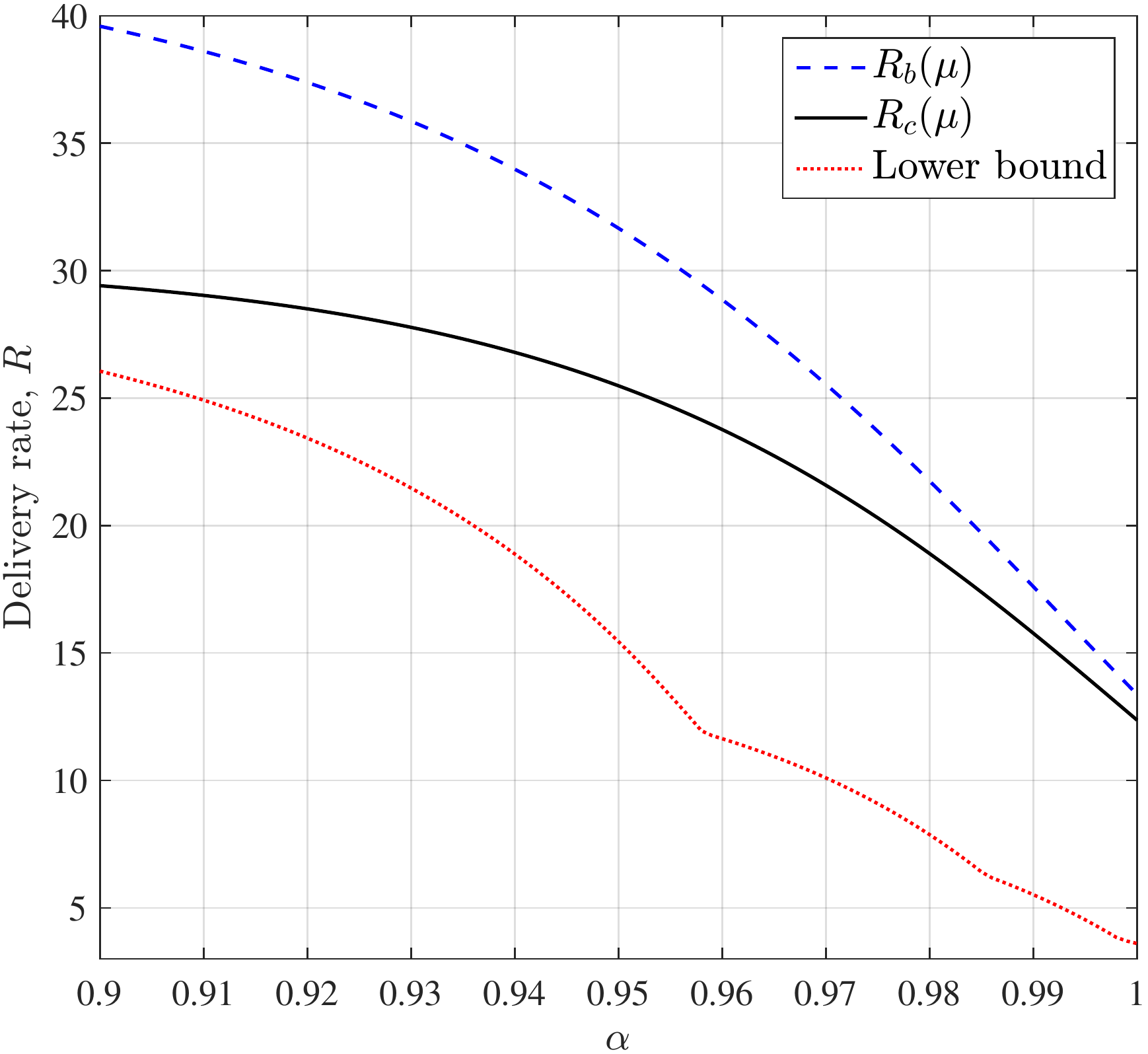}
\caption{Delivery rate versus $\alpha \in \left[ 0.9, 1 \right]$, where $M_k = {\alpha ^{K - k}}M$, $N=30$, $K=45$, and $M=2$.} 
\label{N_75_K_90_AlphaVary}
\end{figure}

In order to see the effect of skewness of the cache capacities on the delivery rate, in Fig. \ref{N_75_K_90_AlphaVary}, the delivery rate of different schemes are plotted as a function of $\alpha \in \left[ 0.9, 1 \right]$, for $N=30$, $K=45$, and the largest cache capacity of $M=2$. The delivery rate of the proposed decentralized coded caching scheme is lower than the one presented in \cite{WangHeterogenous} for the whole range of $\alpha$ values under consideration, while the gain is more pronounced for smaller values of $\alpha$, i.e., as the distribution of cache capacities becomes more skewed. We also observe the gap to the cut-set lower bound also diminishes in this regime.

\section{Conclusions}\label{Conc}
In this paper, we have studied coded caching to users with distinct cache capacities, and proposed a novel decentralized coded caching scheme that improves upon the best known delivery rate in the literature. The improvement is achieved by improving the delivery of bits that have been cached by none of the users, or by only a single user. In particular, the proposed scheme exploits the group-based coded caching scheme we have introduced previously for centralized caching in a system with homogeneous cache capacities \cite{MohammadQianDenizITW}. Our numerical results show that the improvement upon the scheme proposed in \cite{WangHeterogenous} is even more pronounced when the cache capacities of the users are more skewed.

We are currently aiming to improve the delivery rate for larger values of cache capacities by finding a more efficient coded delivery scheme for the delivery of the missing bits of the requested files that have been cached by more than one user. 

\bibliographystyle{IEEEtran}
\bibliography{Report_Asilomar}

\end{document}